\documentclass[%
reprint,
superscriptaddress,
 amsmath,amssymb,
 aps,
floatfix,
]{revtex4-2}

\usepackage{graphicx}
\usepackage{dcolumn}
\usepackage{bm}
\usepackage{siunitx} 
\usepackage{braket}

\usepackage{xcolor}
\usepackage[normalem]{ulem}
\usepackage{soul}
\usepackage[pagebackref=false]{hyperref}

\begin{document}

\author{K.~Hecker}
\thanks{These two authors contributed equally.}
\author{S. M\"oller}
\thanks{These two authors contributed equally.}

\author{H. Dulisch}
\affiliation{JARA-FIT and 2nd Institute of Physics, RWTH Aachen University, 52074 Aachen, Germany,~EU}%
\affiliation{Peter Gr\"unberg Institute  (PGI-9), Forschungszentrum J\"ulich, 52425 J\"ulich,~Germany,~EU}
\author{Ş. Duman}
\affiliation{Institute for Theoretical Physics, TU Wien, 1040 Vienna, Austria, EU}

\author{L. Stecher}
\author{L. Valerius}\thanks{Present address: Department of Physics, Humboldt Universität zu Berlin, 12489 Berlin, Germany}
\author{T.~Deußen}\thanks{Present address: CST – Compound Semiconductor Technology, RWTH Aachen University, 52074 Aachen, Germany}
\author{S. Ravuri}
\affiliation{JARA-FIT and 2nd Institute of Physics, RWTH Aachen University, 52074 Aachen, Germany,~EU}%

\author{K.~Watanabe}
\affiliation{Research Center for Electronic and Optical Materials, National Institute for Materials Science, 1-1 Namiki, Tsukuba 305-0044, Japan}
\author{T.~Taniguchi}
\affiliation{Research Center for Materials Nanoarchitectonics, National Institute for Materials Science, 1-1 Namiki, Tsukuba 305-0044, Japan}%

\author{F. Libisch}
\affiliation{Institute for Theoretical Physics, TU Wien, 1040 Vienna, Austria, EU}
\author{C.~Volk}
\author{C.~Stampfer}
\email{stampfer@physik.rwth-aachen.de}
\affiliation{JARA-FIT and 2nd Institute of Physics, RWTH Aachen University, 52074 Aachen, Germany,~EU}%
\affiliation{Peter Gr\"unberg Institute  (PGI-9), Forschungszentrum J\"ulich, 52425 J\"ulich,~Germany,~EU}%

\title{Radio-frequency charge detection on graphene electron-hole double quantum dots}
\date{\today}

\begin{abstract}
High-fidelity detection of charge transitions in quantum dots (QDs) is a key ingredient in solid state quantum computation.
We demonstrate high-bandwidth radio-frequency charge detection in bilayer graphene quantum dots (QDs) using a capacitively coupled quantum point contact (QPC). The device design suppresses screening effects and enables sensitive QPC-based charge readout. The QPC is arranged to maximize the readout contrast between two neighboring, coupled electron and hole QDs.
We apply the readout scheme to a single-particle electron-hole double QD and demonstrate time-resolved detection of charge states as well as magnetic field dependent tunneling rates.
This promises a high-fidelity readout scheme for individual spin and valley states, which is important for the operation of spin, valley or spin-valley qubits in bilayer graphene. 
\end{abstract}
\maketitle

A fast and state-selective readout method is a prerequisite for operating semiconductor quantum dots (QDs) as hosts for qubits~\cite{Burkard2023Jun}. 
It is a well-established technique to use capacitively coupled quantum point contacts (QPCs) or single-electron transistors to read out charge transitions in semiconductor QDs~\cite{Chatterjee2021Mar,Vigneau2023Jun}.
Typically, the conductance of the charge detector is either measured directly by transport~\cite{Elzerman2003Apr,Ihn2009Sep} or indirectly via the reflectance of an impedance matched LC resonant circuit connected to the charge detector~\cite{Schoelkopf1998May,Reilly2007Oct} to reach a higher bandwidth~\cite{Cassidy2007Nov}. 
This technique enables time-resolved single-shot detection of charge and spin states in current semiconductor devices~\cite{Reilly2007Oct,Elzerman2004Jul,Volk2019Aug,Liu2021Jul,Noiri2022Jan,Philips2022Sep,Mills2022Apr,Madzik2022Jan,Lawrie2023Jun}. 
Graphene and bilayer graphene (BLG), with their naturally low hyperfine interaction and low spin-orbit coupling, offer a promising new platform for spin- and valley-based qubits~\cite{Kane2005Nov,Konschuh2012Mar,Wojtaszek2014Jan,trauzettel2007spin}. 
BLG is of particular interest thanks to its gate voltage tunable band gap~\cite{Icking2022Nov} that allows the formation of QPCs and QDs~\cite{Overweg2018Jan,Banszerus2020May,eich2018spin,banszerus2020electron}.
So far, the energy spectra of single-~\cite{Banszerus2021Sep,Kurzmann2019Jul} and two-particle BLG QD states~\cite{Moller2021Dec} have been investigated and single-particle spin and valley relaxation times have been studied by standard DC transport techniques~\cite{Banszerus2022Jun,Banszerus2025Jul}. 
Blockade mechanisms of spin and valley states have been demonstrated in BLG double QDs (DQDs)~\cite{Tong2021Jan,Tong2024Jan,Moller2025Apr} as well as a particle-hole symmetry protected spin-valley blockade~\cite{Banszerus2023Jun}. 
In addition, charge sensing has been successfully implemented in BLG QD devices, both by transport measurements of a capacitively coupled QD~\cite{kurzmann2019charge} and by dispersive readout techniques~\cite{Banszerus2021Mar2,Ruckriegel2024Jun}.
This allowed for time-resolved detection of charge carriers in QDs~\cite{Garreis2023Jan,Gachter2022May} and DQDs~\cite{Garreis2024Mar} in the Hz to low kHz regime.
Nevertheless, to study real-time dynamics of QDs, charge detection with a high single-shot bandwidth is crucial. 

Here, we present radio-frequency (RF) reflectometry detection of charge states in an electron-hole DQD, both via time-averaged and time-resolved measurements.
The RF reflectometry technique allows to reach a bandwidth of up to several MHz.
The device has been tailored towards maximum readout contrast between the electron and hole QDs by an in-line arrangement of the QPC with the QDs. 
We find that the in-plane capacitive coupling can be improved by reducing the charge carrier density between the QDs and the QPC charge detector.
We identify the effects of screening and the lateral distance between the QDs and the detector on its sensitivity by comparison with a self-consistent Schrödinger-Poisson solver. 
Finally, we perform charge detection in the time domain at the interdot transition of a weakly coupled single-particle electron-hole DQD. 
Thereby, we study the magnetic field dependence of single electron-hole tunneling rates.
Time-resolved charge-detection of single-particle electron-hole DQDs marks an important step towards the readout of spin and valley qubits in BLG~\cite{trauzettel2007spin} or wide band terahertz photon detection~\cite{Bandurin2018Dec}.

\begin{figure*}[p]  
    \centering
    \includegraphics[draft=false,keepaspectratio=true,clip,width=\textwidth]{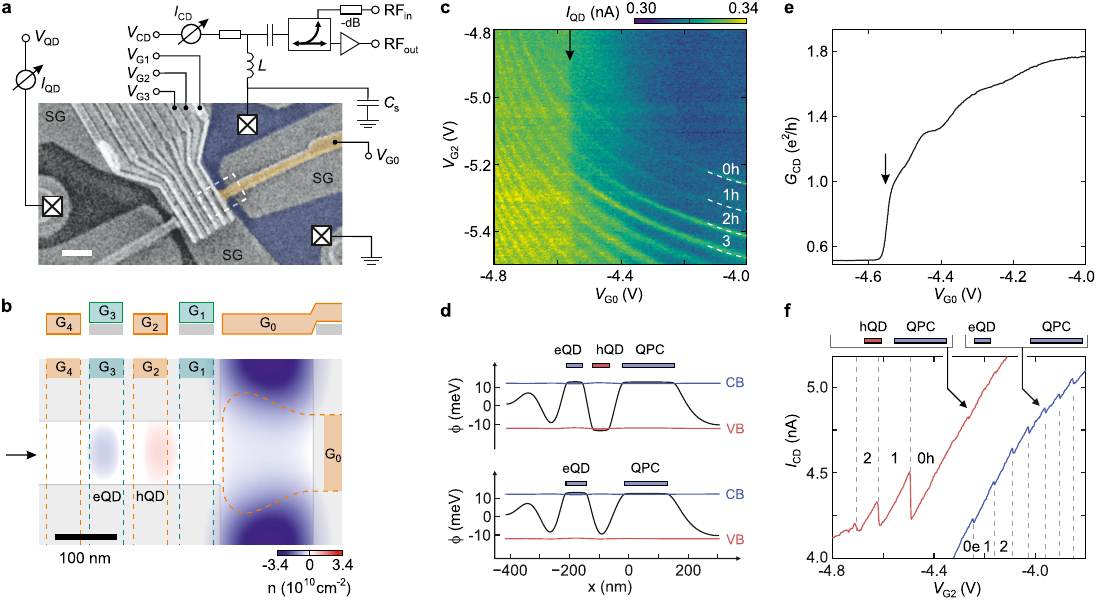}
    \caption{(a) False-colored scanning electron microscopy image of the BLG QD device. Scale bar is \SI{500}{nm}. Split gates (SGs) define conducting channels; finger gates G1--G3 control the potential along the narrow channel, defining and controlling QDs and gate G0 (orange) defines a QPC in the wide channel (blue) acting as charge detector. 
    Via three ohmic contacts ($\boxtimes$) bias voltages can be applied and current through the device can be measured.
    One of the contacts is additionally connected to an inductor, $L=\SI{3.3}{\mu H}$, forming an LC resonant circuit together with the stray capacitance of the bond wires, $C_\mathrm{s}\approx\SI{0.6}{pF}$, matching the impedance of the QPC to the $\SI{50}{\mathrm{\Omega}}$ RF line. For reflectometry measurements, an RF signal ($\mathrm{RF_{in}}$) is applied to the resonant circuit via cryogenic attenuators ($\SI{-36}{dB}$) and a directional coupler ($\SI{-20}{dB}$). The reflected signal is amplified ($\SI{+35}{dB}$, Cosmic Microwave CITLF3) at $\SI{4}{K}$, followed by additional $\SI{+28}{dB}$ at room temperature. A lock-in amplifier (Zurich Instruments UHFLI \SI{600}{MHz}) is used for signal generation and homodyne detection. 
    (b) Schematic of the gate structure together with the calculated charge carrier density $n$ induced in the BLG (scale bar is \SI{100}{nm}). For the calculation the gate voltages were chosen as $V_\mathrm{G1}=\SI{-4.44}{V}$, $V_\mathrm{G2}=\SI{-4.795}{V}$, $V_\mathrm{G3}=\SI{-3.52}{V}$, $V_\mathrm{G4}=\SI{-4.54}{V}$, $V_\mathrm{G0}=\SI{-4.17}{V}$, $V_\mathrm{SG}=\SI{-3.88}{V}$, $V_\mathrm{BG}=\SI{4}{V}$. 
    (c) Current $I_\mathrm{QD}$ as a function of the gate voltages $V_\mathrm{G0}$ and $V_\mathrm{G2}$ with $V_\mathrm{CD} = \SI{250}{\mu V}$, $V_\mathrm{QD} = 0$. Coulomb resonances of a single hole QD formed in the narrow channel close to the T-junction appear. The vertical feature (see arrow) is caused by the formation of a QPC in the wide channel. 
    (d) Potential profiles along the narrow channel (black lines) in between bulk conduction (CB) and valence band (VB) for two different gate voltage configurations defining an electron-hole DQD (upper panel, corresponding to b) and an electron QD (lower panel).
    (e) Conductance $G_\mathrm{CD}$ of the charge detector operated in the QPC regime ($V_\mathrm{CD} = \SI{80}{\mu V}$).
    Here, the two channels were isolated with G1 and a single hole QD is present in the QD channel (see left schematics in panel (f)). The arrow indicates the operating point set in the following experiments. 
    (f) Jumps in $I_\mathrm{CD}$ as a function of $V_\mathrm{G2}$ (see labels and dashed lines) show the formation of single QDs in the few hole ($V_\mathrm{G4} = \SI{-4.3}{V}$, red) and few electron ($V_\mathrm{G4} = \SI{-5.35}{V}$, blue, offset by $\SI{-0.5}{nA}$ for clarity) regimes (see also corresponding schematics on the top)}. \label{f1}
\end{figure*}

The device consists of a heterostructure of BLG encapsulated between two crystals of hexagonal boron nitride (hBN), each $\approx$~\SI{50}{nm} thick, placed on a graphitic back gate (BG). 
Split gates (SGs, see Fig.~\ref{f1}a) are used to form conducting channels. 
In the wide channel (highlighted in blue in Fig.~\ref{f1}a), a QPC beneath the finger gate G0 (orange) acts as a capacitively coupled charge detector.
This channel forms a T-junction with the narrow channel beneath the other finger gates (FGs), where the QDs will be localized. 
This gate architecture is tailored to enhance the capacitive coupling between detector and QDs, in particular to optimize the sensitivity to interdot charge transitions. The relative distance of the detector to adjacent QDs is maximized, following designs of other semiconductor QD devices~\cite{Philips2022Sep}. 
For details on device design and fabrication, we refer to the Supplementary Material and to Ref.~\cite{banszerus2020electron}.

Fig.~\ref{f1}b shows a schematic of the top gates at the T-junction (dashed box in Fig.~\ref{f1}a) together with the simulated charge carrier density in the BLG for an exemplary set of gate voltages to illustrate how the device is operated. 
Here, an electron and a hole QD are formed below the FGs, while a QPC is formed below gate G0 at the T-junction. 
The simulation of a corresponding band edge profile along the center of the narrow channel is shown in the upper panel of Fig.~\ref{f1}d. 
Here, the black curve is the calculated electrostatic potential $\phi$ varying in between the bulk conduction and valence band edges.
Importantly, the charge carrier occupations of the individual QDs can be changed by varying the gate voltages, also allowing to form e.g.~just a single electron QD (lower panel in Fig.~\ref{f1}d) or a single hole QD (see left schematics in Fig.~\ref{f1}f).
The simulation is based on a self-consistent Schrödinger-Poisson solver, where a full 3D model of our device was implemented by an adaptive finite element grid with the NGSolve software package \cite{NGSOLVE} (see Supplementary Material).
The simulations show also that the thickness of the hBN crystals strongly affects the electrostatics of the device. In case of thin hBN, screening by the close metal gates and the graphite back gate strongly limits the sensitivity of the charge detector (see Supplementary Material, Sec.~C).

In the following, the channels between the split gates are set to an n-type polarity by applying a positive voltage to the BG.
First, we apply negative voltages $V_\mathrm{G0}$ and $V_\mathrm{G2}$ to form an n-type QPC in the wide channel and a hole QD in the narrow channel (see QPC and hQD in Fig.~\ref{f1}b).
In Fig.~\ref{f1}c, we investigate the configuration of the QPC and just a single nearby hole QD (see left schematic in Fig.~\ref{f1}f) by measuring the current through the narrow channel, $I_\mathrm{QD}$, as a function of $V_\mathrm{G0}$ and $V_\mathrm{G2}$. 
Coulomb resonances indicate the formation of a hole QD, the occupation (0h, 1h, \ldots) of which can be controlled by $V_\mathrm{G2}$. 
With decreasing voltage $V_\mathrm{G0}$, the hole QD is dragged towards the charge detector, as shown by the curvature of the Coulomb peaks. 
To reduce screening of the capacitive coupling between the charge detector and the QDs, G1 is used to deplete the charge carrier density in between the QPC and the hole QD (compare with Fig.~\ref{f1}b).
This results in a low tunnel coupling, where transport through the QD can no longer be measured directly.
We then measure the conductance $G_\mathrm{CD} = I_\mathrm{CD}/V_\mathrm{CD}$ of the wide channel as a function of $V_\mathrm{G0}$, which is depicted in Fig.~\ref{f1}e, where quantized steps indicate the formation of a QPC. The pinch off of the QPC conductance leads to the change in background current in Fig.~\ref{f1}c (see arrow). 
The residual conductance below the pinch off originates from a parallel conductive channel which is not controlled by the gate voltage $V_\mathrm{G0}$, limiting the maximum resistance to $\SI{50}{k\Omega}$.
To optimize the sensitivity of the charge detector, the QPC is tuned to its steepest flank, highlighted by the arrow in Fig.~\ref{f1}e, where slight changes in the capacitive environment of the QPC translate into strong changes of its resistance. 
By measuring the current $I_\mathrm{CD}$, the charge occupation of the QDs in the narrower channel can be measured indirectly. 
Fig.~\ref{f1}f shows $I_\mathrm{CD}$ as a function of $V_\mathrm{G2}$. By decreasing $V_\mathrm{G2}$, the number of holes occupying the QD under G2 increases, starting from an empty hole QD (red trace in Fig.~\ref{f1}f). Each charge transition causes a sharp jump in $I_{\mathrm{CD}}$.
When applying a more negative voltage of $V_{\mathrm{G4}}=\SI{-5.35}{V}$, a barrier is formed in the channel, which leads to a single electron QD emerging below G3 (see lower panel in Fig.~\ref{f1}d and right schematic in Fig.~\ref{f1}f).
With decreasing $V_{\mathrm{G2}}$, this electron QD is emptied, as shown by jumps in $I_{\mathrm{CD}}$ (blue trace in Fig.~\ref{f1}f).
Here, the area below G2 acts as the barrier of the electron QD.
In this case, the step height of the charge detector response is less pronounced due to the weaker capacitive coupling caused by the larger distance between the electron QD and the QPC charge detector. 

\begin{figure}[!tbh] 
    \includegraphics[width=\columnwidth]{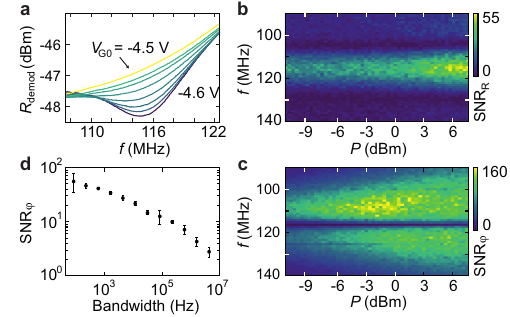}
    \caption{(a) Magnitude of the RF signal as a function of the RF carrier frequency $f$ for different voltages $V_\mathrm{G0}$ applied to the detector gate  (excitation power $P = \SI{-10}{dBm}$). 
    At the resonance frequency $f_\mathrm{res}$ of the tank circuit, the reflected signal is minimized. 
    (b) Signal-to-noise ratio (SNR) of the magnitude $R_\mathrm{demod}$ of the RF signal reflected off the charge detector, measured as a function of $f$ and $P$. The SNR is determined as the ratio of the step height caused by the first charge carrier in the hole QD (0h-1h) and the RMS of the noise. 
    (c) SNR of the phase $\varphi$ of the RF signal. 
    (d) SNR of the phase response as a function of the measurement bandwidth set by the digital first order low pass filter of the lock-in amplifier ($P = \SI{-11}{dBm}$). 
    }
    \label{f2}
\end{figure}

To improve on the bandwidth and the signal-to-noise ratio (SNR) of the detection signal, we use a radio frequency reflectometry technique. Therefore, we excite the resonant circuit and measure the reflected signal in a homodyne demodulation scheme.
Fig.~\ref{f2}a shows the magnitude $R_\mathrm{demod}$ of the demodulated signal as a function of the excitation frequency $f$ for voltages applied to the detector gate between $V_\mathrm{G0}=\SI{-4.6}{V}$ and \SI{-4.5}{V}. 
The total resistance of the charge detector, including the channel resistance and contact resistances, does not reach the matching condition of $R_\mathrm{CD}=\SI{100}{k\Omega}$, since it is limited to $\SI{50}{k\Omega}$ (c.f.~Fig.~\ref{f1}e).  
However, a change of $\SI{1.5}{dBm}$ can still be observed at resonance ($f_\mathrm{res}=\SI{115}{MHz}$). 
To quantify the performance of the charge detector using the RF reflectometry technique, we focus on its response to the first hole being loaded to the hole QD (see Fig.~\ref{f1}f, transition 0h-1h). 
The hole QD causes a stronger response in the CD signal than the electron QD because of its smaller lateral distance to the CD. Thus, the SNR provides an upper bound for the CD response to QDs that are farther away.
We investigate the SNR, which is defined by $\mathrm{SNR_R} = \delta R_\mathrm{step}/\langle R \rangle$ and $\mathrm{SNR}_\varphi = \delta \varphi_\mathrm{step}/\langle\varphi\rangle$, where $\delta R_\mathrm{step}$ and $\delta \varphi_\mathrm{step}$ are the height of the step induced by the addition of a single charge in the demodulated amplitude ($R$) or phase ($\varphi$).
$\langle R \rangle$  and $\langle\varphi\rangle$ are the standard deviation of the signal.
Fig.~\ref{f2}b shows $\mathrm{SNR_R}$ as a function of the excitation frequency $f$ and power $P$. As expected, $\mathrm{SNR_R}$ reaches its maximum at resonance and increases with $P$. 
The SNR of the phase, $\mathrm{SNR}_\varphi$, shows an off-resonance maximum of 160 for $P$ in the range of $-3$ to \SI{0}{dBm} (see Fig.~\ref{f2}c). 
This is three times larger than the maximum $\mathrm{SNR_R}$ achieved at a larger excitation, $P=\SI{6}{dBm}$. 
We attribute the reduced $\mathrm{SNR}_\varphi$ at $P>\SI{0}{dBm}$ to power broadening of the phase response.
At high frequencies, $\mathrm{SNR}_\varphi$ scales with the bandwidth as expected in the presence of white noise, $\mathrm{SNR}_\varphi \propto 1/\sqrt{\mathrm{BW}}$, indicating an SNR above unity up to a bandwidth above \SI{10}{MHz} (see Fig.~\ref{f2}d). 
However, the current setup is limited by the sampling rate of the used RF lock-in amplifier to \SI{7}{MHz}.

\begin{figure}[!thb]
    \centering
    \includegraphics[draft=false,keepaspectratio=true,clip,width=\linewidth]{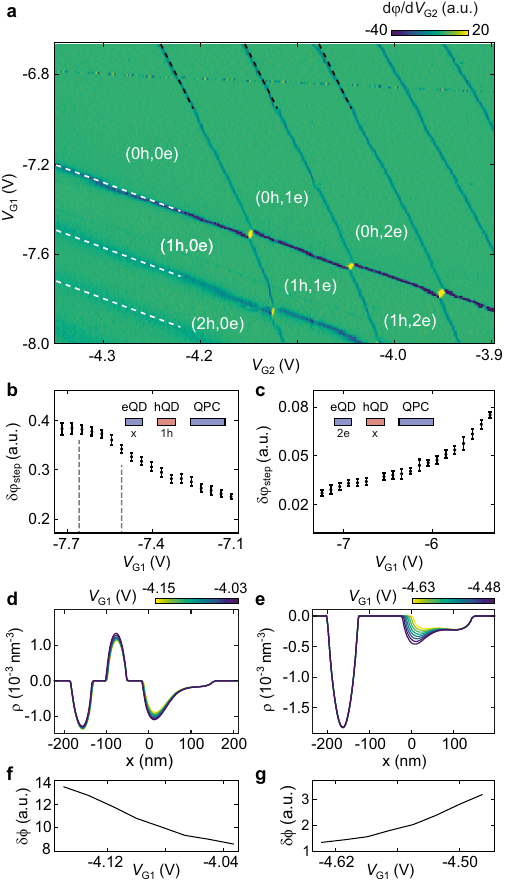}
    \caption{(a) Charge stability diagram of the differential phase $\mathrm{d} \varphi/\mathrm{d}V_\mathrm{G2}$ of the reflected RF signal as a function of the gate voltages $V_\mathrm{G1}$ and $V_\mathrm{G2}$. A hole (electron) QD is defined under gate G1 (G2). The charge carrier occupation of electron (e) and hole (h) QD (see labels) can be tuned down to (0h,0e). 
    (b-c) Step height $\delta\varphi_{\mathrm{step}}$ of the charge detector response at the (b) 0h-1h, (c) 1e-2e charge transition (dashed lines in (a)) as a function of $V_\mathrm{G1}$. Charge transitions of the electron QD occur at the gate voltages indicated by dashed lines. The inset depicts the corresponding charge configuration. 
   (d-e) Line cut of the charge carrier density $\rho$ along $y=0$ in Fig.~\ref{f1}b calculated for different $V_\mathrm{G1}$. The gate voltages $V_\mathrm{G2}$ -- $V_\mathrm{G4}$ have been chosen such that (d) the hole QD is occupied by roughly 1 hole and (e) the electron QD contains two electrons. 
   (f-g) Average calculated potential difference $\delta\phi$ at the QPC due to a change in occupation of (f) the hole QD (g) the electron QD as a function of $V_\mathrm{G1}$ for the configurations shown in the insets of (b) and (c), respectively. }\label{f3}
\end{figure}

In addition to the power and frequency of the applied RF carrier, the sensitivity of the charge detector is also influenced by the lateral distance of the QDs and the QPC as well as screening effects due to a finite charge carrier density in between them.
To investigate these effects, we measure the response of the charge detector on the electron and hole QD charge states using RF reflectometry, while varying the voltage $V_\mathrm{G1}$ applied to the gate in between the QPC and QDs.
The charge stability diagram in Fig.~\ref{f3}a shows the derivative of the phase of the reflected signal, $d\varphi/dV_\mathrm{G2}$ as a function of $V_\mathrm{G1}$ and $V_\mathrm{G2}$. 
To keep the charge detector at its operating point, we correct for electrostatic cross-talk from gate G1 by varying $V_\mathrm{G0}$ according to $V_\mathrm{G0} = -4.737\,\mathrm{V}-0.0193 \cdot V_\mathrm{G1}$.

The (0h,0e) region in the upper left of the charge stability diagram (Fig.~\ref{f3}a) corresponds to both QDs being empty. Sweeping \(V_\mathrm{G1}\) to more negative values loads holes into the QD between G1 and G2 (Fig.~\ref{f1}b), while increasing \(V_\mathrm{G2}\) adds electrons to the neighboring electron QD (see corresponding labels).
The hole QD exhibits a stronger readout contrast than the electron QD, consistent with its closer lateral proximity to the charge detector.

Note that \(V_\mathrm{G1}\) influences the capacitive coupling between the charge detector and the QDs, and hence the detector sensitivity, via two mechanisms:  
(i) it shifts the lateral position of the QDs, changing their distance to the detector;  
(ii) it tunes the carrier density in the region between detector and QDs, thereby modifying the electrostatic screening.
To study this in more detail, we extract the step height \(\delta\varphi_\mathrm{step}\) as a function of \(V_\mathrm{G1}\) for two distinct charge transitions (white and black dashed lines in Fig.~\ref{f3}a).  
For the \((1\mathrm{h},x\mathrm{e})\) configuration—adding the first hole to the hole QD—we observe that the detector response increases as \(V_\mathrm{G1}\) is made more negative (Fig.~\ref{f3}b).  
In contrast, for the \((x\mathrm{h},2\mathrm{e})\) configuration--transitioning from one to two electrons in the electron QD--the response increases with more positive \(V_\mathrm{G1}\) (Fig.~\ref{f3}c).
This charge transition was selected because of its high signal-to-noise ratio. Note that neighboring transitions show a similar qualitative behavior (see Supplementary Material). 
To understand these opposing trends, we compare the experimental data with Schr\"odinger–Poisson simulations of the charge carrier density.  
The results show that the relative impact of (i) the QPC–QD distance and (ii) screening by intermediate charges depends on the spatial charge distribution.  
For the \((1\mathrm{h},x\mathrm{e})\) case, the simulations reveal a change only in the amount of charge between the QD and the QPC charge detector (Fig.~\ref{f3}d), whereas in the \((x\mathrm{h},2\mathrm{e})\) case the QPC position also shifts significantly (Fig.~\ref{f3}e).  
Consequently, the simulations reproduce the experimentally observed opposite trends in \(\delta\varphi_{\mathrm{step}}\) as a function of \(V_{\mathrm{G1}}\) (Figs.~\ref{f3}f,g).
This agreement between experiment and simulation establishes a consistent picture of how \(V_{\mathrm{G1}}\) controls the detector sensitivity via distance and screening effects.  

\begin{figure}[!thb]
  \includegraphics[width=\columnwidth]{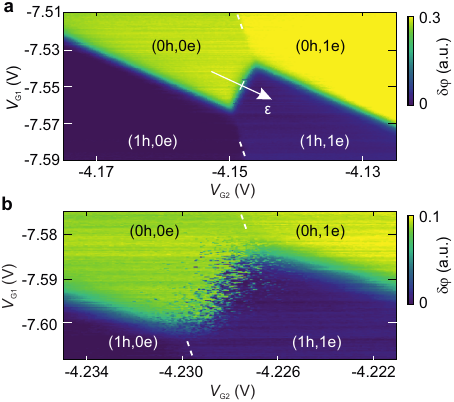}
  \caption{(a) Charge stability diagram showing a close-up of the single-particle electron-hole transition measured in Fig.~\ref{f3}a at $B = \SI{0}{T}$. $\varepsilon$ is the detuning energy between the charge states (0h,0e) and (1h,1e). A linear background caused by capacitive cross talk between the gate voltages $V_\mathrm{G1}$ and $V_\mathrm{G2}$ and the charge detector was subtracted. 
  (b) Same as in (a) but with a perpendicular magnetic field of $B = \SI{1.5}{T}$ applied. }  
  \label{f4}
\end{figure}

In the following, we focus on the single-particle electron–hole regime.  
Fig.~\ref{f4}a shows a close-up of the \((0\mathrm{h},0\mathrm{e}) \leftrightarrow (1\mathrm{h},1\mathrm{e})\) charge transition (see Fig.~\ref{f3}a), corresponding to the creation and annihilation of electron–hole pairs~\cite{Banszerus2023Jun}.  
Here, the different readout contrast for electrons and holes becomes more apparent.  
The in-line arrangement of the QDs leads to strong contrast along the interdot transition (\(\varepsilon = 0\)) and reveals four clearly distinguishable charge states: \((0\mathrm{h},0\mathrm{e})\), \((1\mathrm{h},0\mathrm{e})\), \((0\mathrm{h},1\mathrm{e})\), and \((1\mathrm{h},1\mathrm{e})\).
Fig.~\ref{f4}b shows a similar measurement as in Fig.~\ref{f4}a, but with an out-of-plane magnetic field of $B = \SI{1.5}{T}$ applied.
The finite magnetic field reduces the interdot tunneling rates which is reflected in a random telegraph signal (RTS), close to the degeneracy of the (0h,0e) and (1h,1e) charge states, i.e. at $\varepsilon = 0$, indicating that time-resolved charge detection might be feasible.

\begin{figure*}[!thb]  
    \centering
    \includegraphics[draft=false,keepaspectratio=true,clip,width=\textwidth]{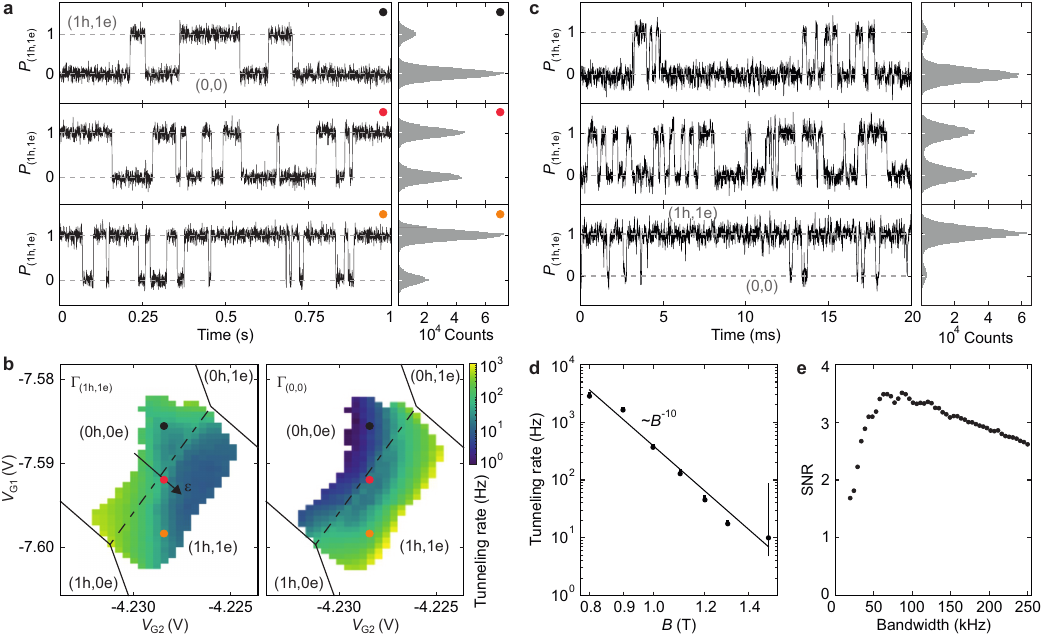}
    \caption{
    (a) Time-resolved detection of (0h,0e) $\leftrightarrow$ (1h,1e) charge transitions at $B = \SI{1.3}{T}$ recorded in a time span of \SI{1}{s} with a sampling rate of \SI{20}{kS/s}. The data has been low pass filtered at \SI{7.5}{kHz}. Histograms of the data sets with a trace length of \SI{50}{s} are shown at the side. 
    (b) Tunneling rates $\Gamma_\mathrm{(1h,1e)}$ (left) and $\Gamma_\mathrm{(0,0)}$ (right) for the annihilation and creation of an electron hole pair, respectively, as a function of $V_\mathrm{G1}$ and $V_\mathrm{G2}$ ($B=\SI{1.3}{T}$). Each data point has been obtained from a single-shot measurement, as in (a). 
    (c) Time-resolved detection in a regime of increased tunnel coupling at $B=\SI{0.9}{T}$ in a time span of \SI{20}{ms} with a sampling rate of \SI{200}{kS/s}. The data has been low pass filtered at \SI{77}{kHz}. Histograms of the data sets with a trace length of \SI{5}{s} are shown at the side.  
    (d) Combined tunneling rate as a function of the perpendicular magnetic field $B$, determined at $\varepsilon = 0$. Fitting a power law yields $\Gamma \propto B^{-10.0\pm0.7}$.
    (e) SNR determined from time-resolved charge detection as a function of the measurement bandwidth ($B = \SI{0.7}{T}$). 
    }
    \label{f5}
\end{figure*}

To realize time-resolved charge detection, we optimized the charge detector for high-bandwidth operation: electrostatic tuning (see discussion above) combined with RF carrier frequency and amplitude optimization increased the sensitivity and enabled bandwidths in the MHz range — an essential requirement for time-resolved QD state readout in qubit operation~\cite{Elzerman2004Jul,Antoniadis2023Jul,Hoehne2014Apr,Noiri2022Jan,Philips2022Sep,Mills2022Apr}.
We assess the performance of the charge detector for frequencies below the RF carrier frequency of around \SI{115}{MHz} by measuring the time-dependent detector signal. Therefore, we reduce the tunneling rates by applying an out-of-plane magnetic field of $B = \SI{1.3}{T}$, similar to Fig.~\ref{f4}b. 
Here, statistical tunnel events can be observed close to the degeneracy of the (0h,0e) and (1h,1e) charge states at $\varepsilon = 0$, hinting towards a low interdot tunneling rate.
To investigate these processes in detail, we measure the demodulated phase $\delta\varphi$ of the reflected signal as a function of time using a digitizer card AlazarTech ATS9440.
Exemplary, Fig.~\ref{f5}a shows \SI{1}{s} long traces out of measurements, which took \SI{50}{s} each. 
The three traces are measured at different $V_\mathrm{G1}$. 
Individual tunneling events can be observed as an RTS between two levels corresponding to the (0h,0e) (lower) and the (1h,1e) (upper) charge configuration. 
The histograms on the right show the distribution of data points recorded over a time span of \SI{50}{s} representing the occupation probability of the two charge configurations. 
At $\varepsilon=0$ (red point in Fig.~\ref{f5}a,b), the DQD is found with equal probability in either of the two configurations. For finite detuning ($|\varepsilon|>0$), the system is either predominantly in the (1h,1e) configuration (orange point) or in the (0h,0e) configuration (black point). The corresponding color-coded positions in the charge stability diagram are shown in Fig.~\ref{f5}b. 
We apply a numerical equivalent of a \textit{Schmitt trigger}~\cite{Horowitz2015} thresholding algorithm to detect single tunneling events and determine the times spans $\tau_\mathrm{(1h,1e)}$ and $\tau_{(0,0)}$ during which the DQD is in the respective state. 
The rates of the charge transitions (1h,1e)$\to$(0h,0e), are $\Gamma_\mathrm{(1h,1e)}=1/\langle \tau_\mathrm{(1h,1e)} \rangle$ and (0h,0e)$\to$(1h,1e), $\Gamma_\mathrm{(0,0)}=1/\langle \tau_{(0,0)}\rangle$, respectively~\cite{Guttinger2011Apr,Garreis2023Jan}. 
Fig.~\ref{f5}b shows the resulting rates $\Gamma_\mathrm{(1h,1e)}$ (left) and $\Gamma_{(0,0)}$ (right) as a function of $V_\mathrm{G1}$ and $V_\mathrm{G2}$ at $B = \SI{1.3}{T}$. At $\varepsilon<0$, $\Gamma_\mathrm{(1h,1e)}$ is increasing while $\Gamma_{(0,0)}$ is decreasing and vice versa for $\varepsilon >0$. This result matches the RTS measurements in Figs.~\ref{f5}a. 
At $\varepsilon=0$, we find equal transition rates of $\Gamma_\mathrm{(1h,1e)}=\Gamma_{(0,0)}\approx \SI{20}{Hz}$. 
Here, resonant tunneling between the electron and hole QD as well as co-tunneling from the reservoirs could contribute to the transition between the charge states~\cite{vanderWiel2002Dec}. 
We decrease the magnetic field to $B = \SI{0.9}{T}$ to increase the interdot coupling. Fig.~\ref{f5}c shows \SI{20}{ms} long time traces that are taken from a measurement recorded for \SI{5}{s} at different positions along the interdot transition (compare with Fig.~\ref{f5}a,b). Here, the tunneling events appear on timescales which are around 100 times faster than at $B=\SI{1.3}{T}$ (see Fig.~\ref{f5}a). To capture these events, the low pass cut-off frequency and the sampling rate of the Alazar card had to be increased to \SI{77}{kHz} and \SI{200}{kS/s}, respectively, leading to a decrease of the SNR.
To quantify the effect of magnetic field tunable tunneling rates, which is in qualitative agreement with previous work~\cite{Tong2022Feb,Banszerus2020Dec,Moller2021Dec,Moller2025Apr}, we measure the transition rates between the charge states (0h,0e) and (1h,1e) at $\varepsilon = 0$ as a function of $B$. 
The result is depicted in Fig.~\ref{f5}d and shows a strong dependency, following a power law of approximately $\Gamma \propto B^{-(10.0\pm0.7)}$. 
By fitting a double Gaussian function to the histograms, which are exemplary shown in Figs.~\ref{f5}a and b, the SNR can be extracted from the time-dependent measurements~\cite{Gustavsson2009Jun}. 
The corresponding data as a function of detection bandwidth is depicted in Fig.~\ref{f5}e (here, exemplary for $B=\SI{0.7}{T}$). We find that a SNR above 2.5 is maintained for a bandwidth exceeding $\SI{250}{kHz}$. 

In summary, we demonstrate RF reflectometry charge detection of an electron-hole DQD in BLG using a capacitively coupled QPC. 
In average signal measurements of the charge detector response, we reach a maximum SNR of 160 and a bandwidth of up to $\SI{7}{MHz}$, which exceeds the bandwidth of 10\,kHz measured with previous two-channel devices~\cite{Kurzmann2019Jul} by three orders of magnitude. So far, comparable values of a bandwidth of up to 10\,MHz were achieved by dispersive sensing using a microwave resonator~\cite{Ruckriegel2024Jun}. 
Improving the matching condition between the resonant circuit and the QPC can further increase the sensitivity of the charge detector. 
The sensitivity can be tuned electrostatically by using the finger gate in between the detector and QDs. 
We identify screening effects and the distance of the detector and QDs as the relevant mechanisms limiting the sensitivity by comparison with a model based on a self-consistent Schrödinger-Poisson solver.
Finally, we demonstrate the distinction of the charge states of a single-particle electron-hole DQD in the time domain with a measurement bandwidth of up to $\SI{250}{kHz}$, exceeding state-of-the-art experiments on electron QDs in BLG, which reach values of 1-10\,kHz~\cite{Garreis2023Jan,Gachter2022May,Garreis2024Mar}. 
With this technique, we determine magnetic field dependent single electron-hole tunneling rates.
Importantly, the electron-hole DQD in BLG is a particle-hole symmetric system with a robust single-particle spin and valley blockade~\cite{Banszerus2023Jun}. 
This mechanism, together with the presented time domain measurements, may allow for high fidelity spin-to-charge and valley-to-charge conversion, providing a reliable read-out scheme for spin and valley states~\cite{trauzettel2007spin}.
This paves the way for measurements of the relaxation and coherence times of spin and valley states, marking a milestone towards the realization of spin, valley and spin-valley qubits.
Beside qubits, further applications of time-resolved readout of electron-hole DQDs in BLG may include wide band single-photon THz detectors~\cite{Bandurin2018Dec,Riccardi2020Jul}, electron-hole pair pumps~\cite{Bordoloi2022Dec} and Cooper pair splitters~\cite{Wang2022Dec}.\\

\textbf{Supporting information}
Details on the device design, on the Schrödinger-Poisson model and on the impact of the insulator thickness supplied as Supporting Information.\\

\textbf{Acknowledgements  }
The authors thank F.~Lentz, S.~Trellenkamp and M.~Otto for help with sample fabrication.
This project has received funding from the Deutsche Forschungsgemeinschaft (DFG, German Research Foundation) under Germany's Excellence Strategy - Cluster of Excellence Matter and Light for Quantum Computing (ML4Q) EXC 2004/1 - 390534769, and by the Helmholtz Nano Facility~\cite{Albrecht2017May}. 
K.W. and T.T. acknowledge support from the JSPS KAKENHI (Grant Numbers 21H05233 and 23H02052), the CREST (JPMJCR24A5), JST and World Premier International Research Center Initiative (WPI), MEXT, Japan.
Ş.D. and F.L. gratefully acknowledge support by the Austrian Science Fund (FWF), doctoral college TU-DX (DOC 142-N, grant DOI 10.55776/DOC142). Electrostatics simulations were performed on the Vienna Scientific Cluster (VSC) 5.\\

\textbf{Author contributions  }
K.H., S.M. C.V. and C.S. conceived this experiment.
K.H, S.M., H.D., S.R., and T.D. fabricated the device. K.H., S.M. and H.D. performed the measurements and analyzed the data with the help of L.V. and L.S.. Ş.D. and F.L. performed the electrostatic simulations of the device architecture. K.W. and  T.T.  synthesized the hBN crystals. C.V. and C.S. supervised the project. K.H., S.M., H.D., C.V., Ş.D., F.L. and C.S. wrote the manuscript with contributions from all authors. K.H. and S.M. contributed equally to this work.\\

\textbf{Data availability}
The data underlying this study are openly available in a Zenodo repository at DOI XXX.\\

\textbf{Competing interests  }
The authors declare no competing interests.\\

\providecommand{\latin}[1]{#1}
\makeatletter
\providecommand{\doi}
  {\begingroup\let\do\@makeother\dospecials
  \catcode`\{=1 \catcode`\}=2 \doi@aux}
\providecommand{\doi@aux}[1]{\endgroup\texttt{#1}}
\makeatother
\providecommand*\mcitethebibliography{\thebibliography}
\csname @ifundefined\endcsname{endmcitethebibliography}  {\let\endmcitethebibliography\endthebibliography}{}

\end{document}


\author{K.~Hecker}
\thanks{These two authors contributed equally.}
\author{S. M\"oller}
\thanks{These two authors contributed equally.}

\author{H. Dulisch}
\affiliation{JARA-FIT and 2nd Institute of Physics, RWTH Aachen University, 52074 Aachen, Germany,~EU}%
\affiliation{Peter Gr\"unberg Institute  (PGI-9), Forschungszentrum J\"ulich, 52425 J\"ulich,~Germany,~EU}
\author{Ş. Duman}
\affiliation{Institute for Theoretical Physics, TU Wien, 1040 Vienna, Austria, EU}

\author{L. Stecher}
\author{L. Valerius}
\author{T.~Deußen}
\author{S. Ravuri}
\affiliation{JARA-FIT and 2nd Institute of Physics, RWTH Aachen University, 52074 Aachen, Germany,~EU}%

\author{K.~Watanabe}
\affiliation{Research Center for Functional Materials, 
National Institute for Materials Science, 1-1 Namiki, Tsukuba 305-0044, Japan}
\author{T.~Taniguchi}
\affiliation{International Center for Materials Nanoarchitectonics, 
National Institute for Materials Science,  1-1 Namiki, Tsukuba 305-0044, Japan}%

\author{F. Libisch}
\affiliation{Institute for Theoretical Physics, TU Wien, 1040 Vienna, Austria, EU}
\author{C.~Volk}
\author{C.~Stampfer}
\email{stampfer@physik.rwth-aachen.de}
\affiliation{JARA-FIT and 2nd Institute of Physics, RWTH Aachen University, 52074 Aachen, Germany,~EU}%
\affiliation{Peter Gr\"unberg Institute  (PGI-9), Forschungszentrum J\"ulich, 52425 J\"ulich,~Germany,~EU}%

\title{Radio-frequency charge detection on graphene electron-hole double quantum dots\\
-- Supplementary Information --}
\date{\today}
\maketitle

\subsection{Device design}
The device consists of a van der Waals heterostructure of BLG encapsulated between two crystals of hexagonal boron nitride (hBN), each approximately \SI{50}{nm} thick, placed on a graphitic back gate (BG). 
Ohmic Cr/Au contacts are etched through the top hBN flake to the BLG. 
Fig.~1a of the Manuscript shows a false color scanning electron microscope (SEM) image of the metallic gate structure fabricated on top of the heterostructure. 
Split gates (SGs) and a graphite BG are used to open a band gap in the BLG, and 
to form conducting channels. 
In the wide channel (highlighted in blue), a QPC beneath the finger gate G0 (orange) acts as a capacitively coupled charge detector.
This channel forms a T-junction with the narrow channel beneath the other finger gates (FGs), where the QDs will be localized.  
A layer of \SI{15}{nm} $\mathrm{Al}_2\mathrm{O}_3$ is placed on top of the SGs, separating them from the Cr/Au FGs 
(orange outlines), which define and control the QDs and the QPC.
These FGs have a center-to-center pitch of \SI{140}{nm} and a width of \SI{70}{nm}. 
The center of the wide channel has a lateral distance of \SI{230}{nm} to the center of FG $\mathrm{G2}$. 
To control the tunnel barriers between the QDs, the QPC charge detector and the reservoirs, an additional layer of FGs (green outlines) is placed on top, separated by another \SI{15}{nm} $\mathrm{Al}_2\mathrm{O}_3$~\cite{Banszerus2021Mar}. For more details on the fabrication of the device, we refer to Ref.~\cite{banszerus2020electron}.

\subsection{Schr\"odinger-Poisson model}
The local charge carrier density $\rho[\phi(\mathbf r'),\partial_z\phi(\mathbf r')](\mathbf r)$ can be simulated by solving Poisson's equation in dielectric media
\begin{equation}
   \nabla(\varepsilon(\mathbf r) \nabla \phi(\mathbf r)) =  \rho[\phi(\mathbf r'),\partial_z\phi(\mathbf r')](\mathbf r),\label{eq:poisson}
\end{equation}
where $\varepsilon(\mathbf r)$ represents the local dielectric constants of the materials, and $\phi(\mathbf r)$ the electrostatic potential. If all permittivities were the same, i.e. $\epsilon(\mathbf r)=const.$, the gradient applied to it would be zero and we would get a single Laplacian on the left-hand side. However, the device consists of materials with different permittivities, hence $\epsilon(\mathbf r)$ changes at interfaces. We account for this by the more general expression above, which can also accommodate tensor-valued $\epsilon$ necessary for different in-plane and out-of-plane permittivities.
The voltages at the different gates serve as Dirichlet boundary conditions for $\phi(\mathbf r)$; at all other boundary points we assume Neumann boundary conditions, $\frac{\partial \phi}{\partial \vec{n}}=0$, where $\vec{n}$ is the normal vector to the surface of the structure at a given boundary point $\vec{r}_{b}$. We assume only BLG carries a charge carrier density, which depends on the local displacement field and its gradient. Consequently, $\rho[\phi(\mathbf r'),\partial_z\phi(\mathbf r')](\mathbf r)$ depends on $\phi(\mathbf{r}')$ for the local shift in potential, and $\partial_z\phi(\mathbf r')$ for the local band gap.
We use a local density approximation, 
\begin{equation}\label{eq:rhor}
\rho[\phi,\partial_z\phi](\mathbf r) \approx \rho\left(\phi(\mathbf r),\partial_z\phi(\mathbf r)\right) \approx -\int_0^{\phi(\mathbf r)}\!\!\!\!\!\!\!\!\!D(\varepsilon';\partial_z \phi(\mathbf r))\mathrm{d}\varepsilon'
\end{equation}
with $D(\varepsilon';\partial_z\phi(\mathbf r))$ being the density of states per unit area of BLG divided by its thickness (to obtain a volume density) at a given energy and displacement field $\partial_z\phi(\mathbf r)$. To simplify further, we assume that the charge carrier density $\rho(x,y,z)=\rho(x,y)$ may only vary in-plane, but not out-of-plane within the $3.35$~Å thickness of BLG. This allows replacing the upper integration bound in Eq.~(\ref{eq:rhor}) $\phi(\mathbf r)$, which varies as a function of three coordinates, by the mean evaluated at the top and bottom height of the BLG sheet at given in-plane coordinates,
\begin{align*}
    \phi(\mathbf{r})\rightarrow \overline\phi(x,y) = 0.5\cdot(\phi(x,y,z_\mathrm{bot.})+\phi(x,y,z_\mathrm{top}))
\end{align*}
\newline
With a self-consistency loop, the charge carrier density in BLG can be numerically calculated for a given set of gate voltages. We start with an initial guess by solving the homogeneous problem, i.e. we set the right-hand side of Eq.~(\ref{eq:poisson}) to zero and obtain $\phi_0$. This first result $\phi_0$ strongly overestimates the induced charge, since no screening is considered. To better condition the start of the self-consistency cycle, we therefore multiply $\phi_0$ by a small number $\gamma \approx 0.025 \ll 1$ while leaving the gradient unchanged. Thus, we evaluate the first induced $\rho_1$ by inserting into Eq.~(\ref{eq:rhor})
\begin{equation*}
    \rho_1=\rho[\gamma\cdot\phi_0, 1\cdot\partial_z\phi_0]
\end{equation*}
This rescaling of $\phi_0$ but not $\partial_z\phi_0$ is only done for the very first iteration to obtain an initial guess for $\rho_1$. For all subsequent iterations we try to find the root of the function
\begin{equation}\label{eq:self_consistency}
    \rho_{n} - \rho[\phi_n, \partial_z\phi_n]=0
\end{equation}
where $\phi_n$ is the solution of Eq.~(\ref{eq:poisson}) with the right-hand side $\rho_n$, which we calculate with NGSolve \cite{NGSOLVE}. At each iteration, the root finder provides a $\rho_n$ for which we solve Poisson's Eq.~(\ref{eq:poisson}), yielding $\phi_n$. The latter we insert into Eq.~(\ref{eq:rhor}) for the charge carrier density, and compare $\rho[\phi_n, \partial_z\phi_n]$ to $\rho_n$. We iterate this procedure until both values (approximately) agree and thus self-consistency is achieved. The difference between the inhomogeneity and the model in Eq.~(\ref{eq:self_consistency}) is computed on a rectangular grid of $N=200\cdot200=40\cdot10^3$ points inside the BLG crystal. Solving the non-linear Poisson's equation~\ref{eq:poisson} amounts to finding a root of a vector-valued function with dimension of $N=40\cdot10^3$. 
\begin{figure}
    \centering
    \includegraphics[width=1\linewidth]{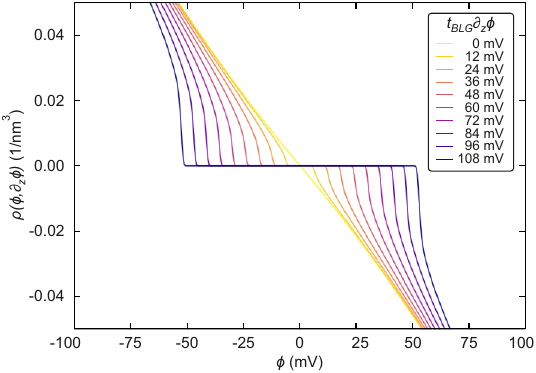}
    \caption{Charge carrier density (Eq.~(\ref{eq:rhor})) as a function of potential $\phi$ for different gradients $\partial_z\phi$. The plateau at $\rho=0$ emerges due to the opening band gap. Note the resulting strong nonlinearity at the band edge. We assume electron-hole symmetry.}
    \label{fig:model}
\end{figure}
\newline
The main challenge of the self-consistency Eq.~(\ref{eq:rhor}) is the band gap in the density of states, the integrand of Eq.~(\ref{eq:rhor}), which changes locally as a function of the out-of-plane derivative $\partial_z\phi(\mathbf{r)}$. As depicted in Fig.~\ref{fig:model}, charge carriers are absent in the band gap. Two kinks show up in the charge carrier density at the Van Hove singularities due to the band edges.
As a consequence, the right-hand side of Eq.~(\ref{eq:rhor}) cannot be linearized locally at every point, thus the tested gradient methods, such as the "good" and "bad" Broyden methods to obtain the root of function \ref{eq:self_consistency}, fail. However, we find that the derivative free spectral residual method \cite{df-sane} does work in finding the root of Eq.~(\ref{eq:self_consistency}). We use the \textit{scipy.optimize.root} implementation \cite{2020SciPy-NMeth}.
\newline
The average potential difference $\delta\phi$ at the QPC caused by a change
in occupation depicted in the simulation curve in the main text (Fig.~3(f)) can be made more precise:
\begin{equation*}
    \delta\phi = \langle\phi\{V_{(i) }\} \rangle_\mathrm{QPC} - \langle\phi\{V_{(i)}+\delta V_{(i) }\} \rangle_\mathrm{QPC} 
\end{equation*}
For this expression, we first calculate the electric potential $\phi$ at a specific configuration (for instance in Fig.~3(e) of the main text we go along a section of the $(1e,0h)$ to $(2e,0h)$ charge transition line in the charge stability diagram) but for different voltages $\{V_{(i)}\}$. Once this line $\{V_{(i)}\}$ is found, we perform another simulation for a small deviation $\delta V_{(i) }$ along this section in the charge-stability diagram. Experimentally, the QPC (and in consequence the measured step height) is sensitive to this change of the electric potential
\begin{equation*}
    \phi_\mathrm{diff}=\phi\{V_{(i) }\} - \phi\{V_{(i)}+\delta V_{(i) }\}
\end{equation*}
We map this function to a single scalar value by calculating the averages of the individual potentials inside BLG under the QPC ($\langle...\rangle_\mathrm{QPC}$). This measure allows us to make general qualitative statements about the change of the magnitude of $\phi_\mathrm{diff}$. However, for a quantitative prediction one also needs to model the bound state spectrum of the BLG Hamiltonian and additionally account for screening due to the integer occupation of electrons and holes. This improvement of the model is computationally very challenging and left for future publications. Nevertheless, since the QPC is far away from the charges, we expect the qualitative trends of Fig.~3 (main text) to hold even in this case. We only expect corrections in the amplitude of $\phi_\mathrm{diff}$ and in the immediate vicinity of the electron and hole bound states, leaving the functional form of the long-range electrostatic field intact.

\begin{table}[h]
    \centering
    \begin{tabular}{|l|l|} \hline
        material & $\epsilon$\\ \hline
        vacuum & 1\\ \hline
        polycarbonate & 3.1\\ \hline
        Al$_2$O$_3$ & 9.1\\ \hline
        hBN$_{\parallel}$ & 4.98\\ \hline
        hBN$_{\perp}$ & 3.03\\ \hline
        BLG$_{\parallel}$ & 1.8\\ \hline
        BLG$_{\perp}$ & 3\\ \hline
    \end{tabular}
    \caption{Scalar and tensor permittivities $\epsilon$ of the materials relevant in the experiment. For hBN and BLG we distinguish between in-plane $\epsilon_\parallel$ and out-of-plane $\epsilon_\perp$ permittivities (see \cite{epsilon_hBN} and \cite{epsilon_graphene}).}
    \label{tab:permittivity}
\end{table}

\begin{figure*}[]
\centering
\includegraphics[draft=false,keepaspectratio=true,clip,width=0.9\linewidth]{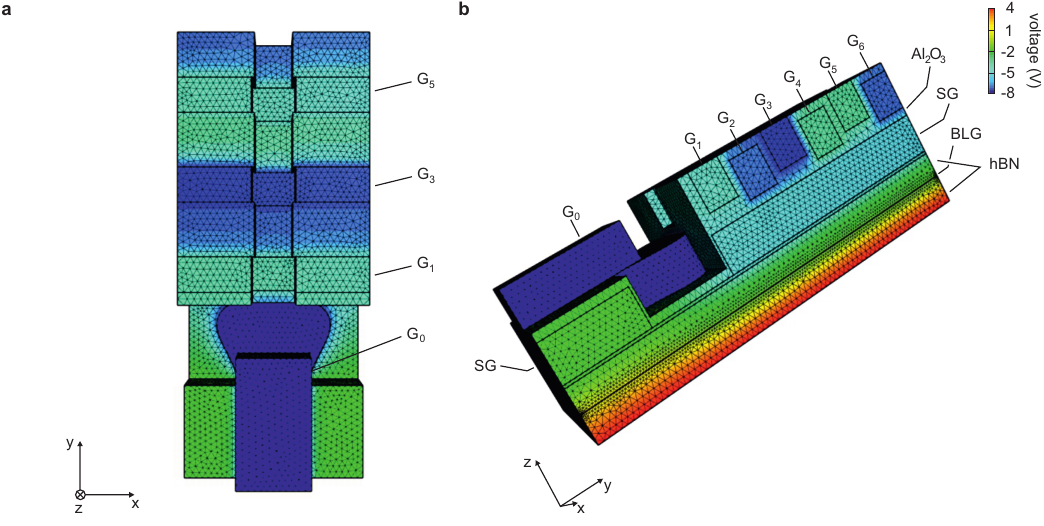}
\caption[Fig00]{(a) Top view and (b) side view of a 3D model of the used device, showing the heterostructure and the gate stack. The color code shows the voltages applied to the gates and the calculated potential throughout the device. A layer of vacuum covering the top gates filling out the simulation box is not shown for visibility.}
\label{fs1}
\end{figure*}

Fig.~\ref{fs1} shows a simulation of the potential $\phi_0$ solving the homogeneous problem for the investigated device geometry and an exemplary set of gate voltages.
Poisson's equation requires permittivities for each material. 
We use the scalar relative permittivities $\epsilon$ as given in Tab.~\ref{tab:permittivity}.

\newpage

\subsection{Impact of the insulator thickness:\\ a toy model}

\begin{figure}[h]
    \centering
    \includegraphics[width=1\linewidth]{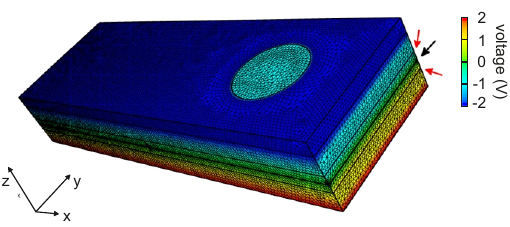}
    \caption{Overview of the toy model. The electric potential $\phi$ which solves the homogeneous problem with charge density $\rho=0$ is plotted for the entire structure, the color bar goes from $-2V$ (blue, top gate voltage $-V_B$) to $2V$ (red, bottom gate voltage $V_B$). The cylinder gate is held at $V_C=-1V$. Neumann boundary conditions apply  $\frac{\partial \phi}{\partial \vec{n}}=0$ to the remaining boundaries. The maximal triangle edge mesh size is chosen to be $2.5~$nm. The hBN thickness depicted here is $20~$nm, both layers of hBN are subdivided into two (the red arrow point to where this happens), such that the region close to the BLG sheet (which the black arrow points to) is resolved better.}
    \label{fig:toy_model_overview}
\end{figure}

The distance between the bilayer and the metal gates strongly impacts the electrostatics of the device. While a thin insulating layer allows for a larger gate lever arm, it also increases screening, leading to much quicker decay of charge-induced potential variations with distance, making charge detection more difficult. Conversely, a thick insulating layer necessitates larger applied voltages, and thus potentially short-circuiting.
\newline
To understand how the insulator thickness affects the
change of electrostatic potential $\phi$ inside the BLG sheet we numerically investigate a toy model. We leave an analytical calculation to a further study.
We focus here on elucidating  the connection between the sample geometry and the resulting electrostatic properties. We choose a simple model of a cylindrical gate on top of a bilayer sheet (see Fig.~\ref{fig:toy_model_overview}) to disentangle the different contributions to the electrostatic potential. Consider a BLG sheet of $300~$nm$\cross100~$nm area sandwiched between two layers of hBN of equal thickness $w$. Below the hBN-bilayer-hBN stack, there is a single bottom gate at voltage $V_B$, and on top there are two gates: a planar gate with a hole at a voltage of $-V_B$, and a cylindrical gate within this hole at voltage $V_C$. 

To understand the influence of the hBN width $w$ we consider the change from one to two electrons inside the dot, i.e., $Q_1=1e=-1$ and $Q_2=2e=-2$ (in atomic units) and fix a displacement field $\delta\phi_{int.lay.}= 30$ meV. These constraints uniquely determine $V_B(w)$, $V_{C,1}(w)$ and $V_{C,2}(w)$: for a given hBN thickness $w$, we determine (i) the $V_B(w)$ needed to arrive at $\delta\phi_{int.lay.}$ within the bulk BLG, evaluated at the left edge, far away from the circular gate. and (ii) the  voltages $V_{C,1}(w), V_{C,2}(w)$ yielding the charges $Q_1$ and $Q_2$ below the circular gate. To obtain $\delta\phi_{12}$ numerically, we solve for $\phi_1$ and $\phi_2$ individually and then take the difference $\delta\phi_{12}=\phi_2-\phi_1$. The non-linear source term $\rho[\phi,\partial_z\phi]$ is given by equation \ref{eq:rhor}, just like the main text. 

\begin{align*}
    \delta\phi_{12}(\mathbf r) &= \phi_1(\mathbf r) - \phi_2(\mathbf r) \\ 
    \nabla(\varepsilon(\mathbf r) \nabla \phi_1(\mathbf r)) &=  \rho[\phi_1,\partial_z\phi_1] := \rho_1(\mathbf r) \\
    \nabla(\varepsilon(\mathbf r) \nabla \phi_2(\mathbf r)) &=  \rho[\phi_2,\partial_z\phi_2] := \rho_2(\mathbf r) \\
    Q_1 &= \int d\vec{r}\rho_1(\mathbf r)  = const. \\
    Q_2 &= \int d\vec{r}\rho_2(\mathbf r)  = const.
\end{align*}

\begin{figure}
    \centering
    \includegraphics[width=1\linewidth]{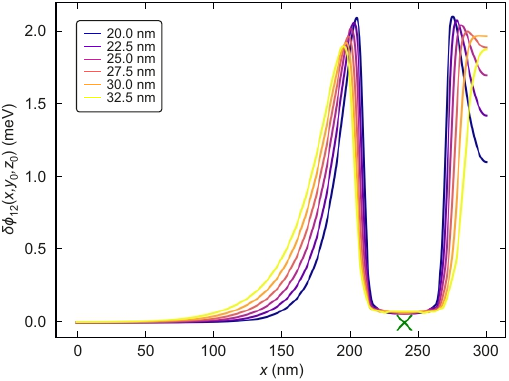}
    \caption{$\delta\phi_{12}$ along the center line $y_0=50~$nm and $z_0=w+0.5\cdot t_{BLG}$ for different hBN thicknesses $w$, where $t_{BLG}$ is the BLG thickness. Exponential tails are elongated with increasing $w$. The center point of the cylindrical gate stands above the green cross. }
    \label{fig:toy_model_phi3_cut}
\end{figure}

\begin{figure}
    \centering
    \includegraphics[width=1\linewidth]{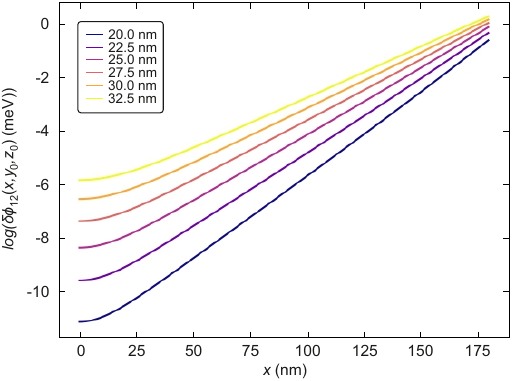}
    \caption{Logarithmic plot of $\delta\phi_{12}$ after all exponential tails emerge at $175~$nm. The slopes are the decay of each exponential. Toward $x=0$ we see flattening of the logarithmic curves and thus flattening of $\delta\phi_{12}$, which is imposed by the Neumann boundary conditions.}
    \label{fig:toy_model_phi3_cut_log_plot}
\end{figure}

We consider hBN thicknesses $w = 20\ldots32.5~$ nm in steps of $2.5~$nm. 
We find that the potential difference $\delta\phi_{12}$  exponentially decreases away from the source (see Figs.~\ref{fig:toy_model_phi3_cut} and \ref{fig:toy_model_phi3_cut_log_plot}), as a function of in-plane distance $r$ in the BLG sheet, where the decay is decreasing with larger values of $w$. Intuitively, this occurs because the planar gates move away with increasing $w$, reducing their importance compared to the charge transition occuring in the BLG sheet.

Our numerical results suggest that, while the equations governing the quantum point contact (QPC) might be complicated, they result in exponential decay of the response, with the decay constant clearly decreasing with increasing $w$. Thus, increasing the insulator height $w$ improves the signal, preventing the field lines emanating from the localized charge from being attenuated. These considerations should intuitively be true very far away from the source, where we approximately have a plate capacitor. Our numerical results remarkably predict that this exponential behavior also holds close to the gates. 

While we performed these calculations for this simple toy model with a single source $\rho_{1,2}$, our reasoning is quite general and even holds if there is some common charge distribution unchanged by the variation of $V_{C,1}\rightarrow V_{C,2}$, such as a $\rho_{\mathrm{left}}$ (potentially acting as a charge detector) on the left side of the structure, as long as this is included in both charge distributions $\rho_1$ and $\rho_2$. From an experimental point of view, a larger $w$ requires larger gate voltages, ultimately limiting the thickness $w$ from above. We are currently working on a more detailed discussion and comparison with an analytical model. 

One of the advantages of the T-shaped model of the device shown in the main text is that it is not an approximate plate capacitor, which is roughly the case in the usual double-channel geometry in other experiments. Thus, for the T-shaped geometry, there is no large top side gate between the QD and the charge detector, which according to our intuition obtained from the toy model of this section, should improve the detector signal $\delta\phi_{1,2}$. 

While the model presented here solves the Poisson equation self-consistently, the quantum-mechanical part is based on the bulk density of states instead of a full solution of the Schr\"odinger (or Dirac) equation. A comprehensive treatment of this problem would need to simultaneously solve Poisson's equation and the quantum mechanical wave equation for the eigenstates of the electrons in the device to account for bound states, which poses a significant computational challenge. We conjecture here that the  evident exponential decay of $\delta \phi_{12}$ with distance suggests that even for such an improved treatment of the quantum mechanical part, the plate capacitor solution of the left-side will determine the long-range electrostatic field. Thus, while quantitative  predictions for, e.g., the height of the exponential tail and its decay constant may depend on the model employed, the qualitative long-range trend should remain unaffected. Ultimately, it is this long-range behavior which determines the sensitivity of the charge detector.

\begin{figure*}[ht!]  
    \centering
    \includegraphics[draft=False,keepaspectratio=true,clip,width=0.65\linewidth]{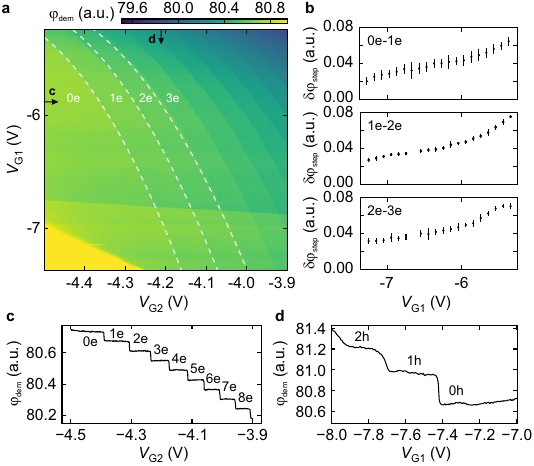}
    \caption{(a) Charge stability diagram showing the phase $\varphi_\mathrm{dem}$ of the reflected RF signal as a function of the gate voltages $V_\mathrm{G1}$ and $V_\mathrm{G2}$. 
    (b) Step heights of the charge transitions 0e-1e, 1e-2e, and 2e-3e extracted from panel (a) (see white dashed lines). 
    (c,d) Line cuts along the $V_\mathrm{G2}$-axis and $V_\mathrm{G1}$-axis (see arrows in panel~(a)) showing the sequential filling of the electron QD. }
    \label{fig:electronResonances}
\end{figure*}

\subsection{Complementary data}
\vspace*{-1em}

Fig.~\ref{fig:electronResonances}a shows an extended range of the charge stability diagram in Fig.~3a of the manuscript.
Multiple electron transitions are visible (see labels). 
The extracted step heights of the transitions 0e-1e, 1e-2e, and 2e-3e (dashed lines) are depicted in Fig.~\ref{fig:electronResonances}b. As can be seen, the qualitative behavior is equal for all three electron resonances. 
For the analysis shown in Fig.~3b and c of the manuscript, we picked the charge transitions 0h-1h and 1e-2e to present results with the best signal-to-noise ratio.
Figs.~\ref{fig:electronResonances}c and d show cuts through the charge stability diagram along constant $V_\mathrm{G1}$ and $V_\mathrm{G2}$ showing sharp steps indicating the sequential filling of the electron and hole QD.

%